\documentclass[fleqn]{POWFI-2018}
\usepackage{graphicx}
\usepackage{subfigure}
\usepackage{amssymb,amsmath}
\usepackage{url}
\begin{document}

\ensubject{subject}

\ArticleType{Article}
%\SpecialTopic{SPECIAL TOPIC: Research Highlights}
\Year{2019}
\Month{January}
\Vol{60}
\No{1}
\DOI{ }
\ArtNo{000000}
\ReceiveDate{, 2019}
\AcceptDate{, 2019}

\newcommand{\PRcal}{J0348$-$2749}
\newcommand{\GRB}{GRB 190114C}
\newcommand{\degr}{$^\circ$}
\newcommand{\arcsec}{$^{\prime \prime}$}
\newcommand{\arcmin}{$^{\prime}$}

\newcommand{\adv}{    Adv. Spa. Res.}
\newcommand{\aap}{    A\&A}
\newcommand{\aaps}{   Astron. Astrophys. Suppl.}
\newcommand{\aapr}{  Astron. Astrophys. Rev.}
\newcommand{\aj}{     AJ}
\newcommand{\apj}{    ApJ}
\newcommand{\apjl}{   ApJL}
\newcommand{\apjs}{   ApJS}
\newcommand{\apss}{   ApSS}
\newcommand{\cjaa}{   Chinese J. Astron. Astrophys.}
\newcommand{\mnras}{  MNRAS}
\newcommand{\nat}{    Nature}
\newcommand{\na}{     New Astron.}
\newcommand{\pasa}{   PASA}
\newcommand{\pasp}{   PASP}
\newcommand{\pasj}{   PASJ}
\newcommand{\sovast}{ {\it Sov. Astronom.}}
\newcommand{\ssr}{    Space Sci. Rev.}
\newcommand{\rmp}{    {\it Rev. Mod. Phys.}}
\newcommand{\physrep}{   Physics Reports}
\newcommand{\araa}{    ARA\&A}
\newcommand{\memras}{    {\it Memoirs of the Royal Astronomical Society}}
\newcommand{\memsai}{ MemSAI}
\newcommand{\nar}{ New Astron. Rev.}

% comment shortcuts
\newcommand{\om}[1]{\textcolor{green}{#1}}

\title{\bf East Asia VLBI Network observations of the TeV Gamma-Ray Burst 190114C} 

\author[1,2]{Tao An}{antao@shao.ac.cn}
\author[3]{Om Sharan Salafia}{}
\author[1,2]{Yingkang Zhang}{}%
\author[3]{Giancarlo Ghirlanda}{}
\author[4,5]{Giovannini Giovannini}{}%
\author[4]{\\ Marcello Giroletti}{}%
\author[6]{Kazuhiro Hada}{}
\author[4,5]{Giulia Migliori}{}
\author[4]{Monica Orienti}{}
\author[7]{Bong Won Sohn}{}

\AuthorMark{An T}

\AuthorCitation{An et al}

\address[1]{Shanghai Astronomical Observatory, Key laboratory of Radio Astronomy, CAS, Nandan Road 80, Shanghai 200030, China}
\address[2]{University of Chinese Academy of Sciences, 19A Yuquan Road, Shijingshan District, 100049 Beijing, China}
\address[3]{Istituto Nazionale di Astrofisica, Osservatorio Astronomico di Brera, Via E. Bianchi 46, I-23807 Merate, Italy}
\address[4]{Istituto Nazionale di Astrofisica, Istituto di Radioastronomia, via Gobetti 101, I40129, Bologna, Italia}
\address[5]{University of Bologna, Department of Physics and Astronomy, via Gobetti 83, I40129, Bologna, Italia}
\address[6]{National Astronomical Observatory of Japan, 2-21-1 Osawa, Mitaka, Tokyo 181-8588, Japan}
\address[7]{Korea Astronomy and Space Science Institute, Yuseong-gu, Daejeon 34055, Korea}

\abstract{
Observations of gamma-ray bursts (GRBs) at Very High Energy (VHE) offer a unique opportunity to investigate particle acceleration processes, magnetic fields and radiation fields in these events.
Very Long Baseline Interferometry (VLBI) observations have 
been proven to be a powerful tool providing unique information on 
the source size of the GRBs at mas scales, as well as their accurate positions and possible  
expansion speeds.
This paper reports on the follow-up observations of \GRB, the first ever GRB detected with high significance at TeV photon energies  
by the MAGIC  telescope, conducted with the East Asia VLBI Network (EAVN) at 22 GHz on three epochs, corresponding to 6, 15 and 32 days after the burst.
The derived maps
do not show any significant source above $5 \sigma$.  
The inferred upper limits on the 
\GRB flux density 
at 22 GHz are used here to constrain the allowable two--dimensional parameter space for the afterglow emission. We find that our limits are consistent with most afterglow parameter combinations proposed so far in the literature.  
This is the first effort for the EAVN to search and monitor a radio transient 
in the Target of Opportunity mode. In addition to the useful constraints on \GRB\ radio emission, experience gained from these 
observations is very helpful for future routine operation of EAVN transient program.
}
\keywords{techniques: interferometric --- galaxies: active --- gamma rays: bursts --- individual: GRB 190114C}

\PACS{}

\maketitle

% \begin{multicols}{2}

%
%________________________________________________ sections below
%
\section{Introduction}           %% first-level sections will be auto-capitalized
\label{sect:intro}

Long duration gamma-ray bursts (GRBs) signpost the death of massive stars and the formation of stellar mass black holes (BHs). 
GRB emission extends from the $\gamma$-rays (the short-lived prompt emission) through the X-rays and optical/near infrared to the radio bands (the long-lived afterglow emission, reviewed in previous works \cite{1995ARA&A..33..415F,2000ARA&A..38..379V,2005AIPC..784..164P}). 
Very high energy (VHE, $E \gtrsim 100 $ GeV) emission from GRB afterglows has been long predicted (e.g., \cite {1999ApJ...512..699C,2001ApJ...559..110Z,2008FrPhC...3..306F}).  VHE detections provide rich information on GRB shock physics and on the conditions of the emitting zone \cite{Derishev2019}. In particular, the detection of the inverse Compton emission component would constrain the characteristic energy of the shock accelerated electrons responsible for the synchrotron emission and the fraction of the outflow kinetic energy distributed among relativistic particles and, through its amplification, magnetic field. On the other hand, the detection of VHE emission from GRBs is very challenging due to the attenuation by extra-galactic background light and/or by internal absorption. 
Despite that hundreds of GRBs have been detected with an afterglow, no emission at tera electron Volt (TeV) photon energy has ever been detected above $3\sigma$ \cite{2017ApJ...846..152V,2018IJMPD..2742003N} until only recently.
On 14 January 2019 at UT 20:57:03, the Burst Alert Telescope (BAT) onboard NASA's Neil Gehrels {\it Swift} Observatory (\textit{Swift} hereafter) detected a bright gamma-ray burst (\GRB) \cite{GCN-Swift}. After 50 seconds, this GRB was detected by the MAGIC telescope at very high energy ($>$300 GeV) \cite{GCN23701} with a significance of $>20\, \sigma$. The burst has been also detected by the Gamma-Ray Burst Monitor (GBM)
\cite{GCN23707} and the Large Area Telescope (LAT) \cite{GCN23709} onboard the \textit{Fermi} satellite. 
Prompt follow-up by {\it Swift} revealed a bright X-ray afterglow and initiated a series of multi-band follow-up observations ranging from X-ray to radio bands.

Other than being the first GRB ever to be detected by ground-based Cherenkov telescopes, \GRB{} is a notable target of particular attraction for a number of other reasons. First, \GRB{} shows evidence of a very early afterglow onset \cite{2019A&A...626A..12R}, allowing for an estimate of its initial bulk Lorentz factor (see also \cite{2019arXiv190910605A}). Second, \GRB{}  was observed and detected across 16 orders of magnitude in frequency, from $\gamma$-rays to radio, within several hours after the burst (see above paragraph), allowing for a study of the multi-band evolution of the afterglow starting early after the trigger
(e.g., for a notable example of how important broadband afterglow modelling can be for the interpretation of this kind of events \cite{2008Natur.455..183R}).
Third, \GRB{} is one of the few GRBs with an early radio detection: 3.2 hr after the burst it was observed by ALMA at 90 GHz, revealing a fading mm-wavelength counterpart \cite{GCN23728}; 4.2 hr after the burst by the VLA at a central frequency of 33.5 GHz \cite{GCN23726} with a flux density of 3.1 mJy at the position of RA (J2000) = 03h38m01s.191, Dec (J2000) = $-$26$^\circ$56\arcmin46\arcsec.72; later ATCA detected the source at 5.5, 9, 17 and 19 GHz with a flux density of $\sim$2 mJy about 1.5 days after the burst \cite{GCN23745}.
A detection was also reported by the RT-22 telescope, with a flux density of $\sim$4 mJy at 36.8 GHz on 2019 January 15 and 16 \cite{GCN23750}.

Fundamental information about GRB jets can be derived from radio observations of their afterglow, including jet energy,  ISM density, and shock parameters \cite{2014PASA...31....8G}.
Very Long Baseline Interferometry (VLBI), owing to its unparalleled high resolution, has proved to be capable of providing unique information on
the shock energy (and its distribution) and expansion velocity.
One such case was the famous and exceptionally bright GRB 030329. 
Very Long Baseline Array (VLBA)  and global VLBI observations determined the accurate size and source expansion velocity of GRB 030329 \cite{2004ApJ...609L...1T,2005ASPC..340..298T,2007ApJ...664..411P,2012ApJ...759....4M}.
Later on, VLBI observations of GRB 170817A (the electromagnetic counterpart of the gravitational wave event GW 170817) revealed a proper motion of the radio centroid and secured constraints on its linear size at milliarcsecond scales, invaluable in telling competing models for the ejecta dynamics apart \cite{2018Natur.554..207M,2018Natur.561..355M,2019Sci...363..968G}. 
Modelling the temporal evolution of radio flux densities offers critical constraints on the progenitor environment.
The multi-band light curves of the extremely bright GRB 130427A can be described by a two-jet model, which favors a low-density progenitor environment  \cite{2014MNRAS.444.3151V,Perley2014,2014Sci...343...48M}.

Multi-epoch European VLBI Network (EVN) and VLBA observations of GRB 151027A played a pivotal role in favouring the wind density profile model, typical of the medium surrounding a massive star in the final stages of its evolution, rather than a homogeneous profile typical of the interstellar medium \cite{Nappo2017}.
Given the exceptional brightness and energy spectrum of GRB 190114C, this seemed another ideal case to take advantage of high angular resolution 
and sensitivity 
provided by the VLBI technique.

Owing to the above reasons, we applied for
VLBI observations of \GRB{} with the 
East Asia VLBI Network (EAVN)
\cite{2018NatAs...2..118A}. It is critical to have a proper characterisation of the emission in the first days, when it is possible to observe the source at high frequency. Moreover, multi-epoch VLBI radio observations starting in the immediate aftermath of the burst enable to determine or constrain the source expansion. In order to 
fulfill
these described goals, we conducted VLBI observations at 22 GHz in three epochs within about one month after the burst.
The VLBI observational results and insights from afterglow modelling  
are presented in this paper.

\section{Methods}
\label{sect:Obs}

We carried out high-resolution follow-up observations of \GRB{} with the East Asia VLBI Network on three epochs of 2019 January 21, 30, and 2019 February 16, which correspond to 6, 15 and 32 days respectively  after the burst discovered on 2019 January 15 (see report in \cite{GCN-EAVN}).
These observations made use of the Director's Discretionary Time (DDT).
The first observation was scheduled as soon as possible, within 1 week after the burst, with the initial purpose of constraining the self-absorption frequency. 
The primary purpose of the second and third epochs was 
to monitor the flux density evolution in order to track the possible evolution of the self-absorption frequency 
and, if possible, to determine or constrain the source expansion speed.
They would 
also have the added value of validating and strengthening any positional constraints derived from the first epoch.

Altogether, 
eight telescopes participated in the observations. The detailed observation setup is presented in Table~\ref{tab:obs}.
The same observation setup was used in the three sessions. Although the target is at a relatively low declination, each station could observe it  
at elevation $>20^\circ$ for most of the 
time during the observing runs and with a maximum elevation angle up to 40$^\circ$. An example of the (u,v) coverage is shown in Fig.~\ref{fig:uvcov}.
The observations were carried out in phase referencing mode. The 
nearby bright source  \PRcal{} (1.2 Jy at 22 GHz; 2.5$^\circ$ away from the target on the plane of the sky) was used as the calibrator. The image of \PRcal{} is displayed in Figure \ref{fig:pha-cal}, showing a compact core-jet structure.
We first calibrated the phases of \PRcal, then applied the complex gain solutions derived from \PRcal{} to the target source data by interpolation. 
The \GRB{} data were finally averaged in frequency within each sub-band, but individual sub-bands were kept to minimize bandwidth smearing. Similarly, the data were time-averaged to 2 seconds for imaging.
Additional details on the observation and data reduction are deferred 
to the Appendix.

We compare the upper limits from our radio observations with standard afterglow model predictions. We model the GRB afterglow emission following Granot et al. \cite{Granot2002}, which accounts for synchrotron emission from electrons in a relativistic shock whose hydrodynamical profile is described by the self-similar \cite{Blandford1976} solution. The number density profile of the external medium in which the shock propagates is parametrized as $n(R)\propto R^{-k}$, where the two relevant scenarios in GRBs are $k=0$ (constant density medium, representing a uniform ISM region) or $k=2$ (the expected density profile of the circum-burst medium produced by the winds of the progenitor star, e.g. \cite{Chevalier2000}). A fraction $\epsilon_\mathrm{e}$ of the energy density in the shocked material is assumed to take the form of relativistic electrons (accelerated into a non--thermal energy distribution with index $p$), while a fraction $\epsilon_{\rm B}$ is assumed to be shared by the magnetic field, amplified by small-scale magneto-hydrodynamic instabilities. These parameters suffice to derive the synchrotron emission, i.e., the spectral shape and normalization, at all times for which the underlying assumptions hold.

\section{Results and Discussion}
\label{sect:result}

Figure \ref{fig:GRB} shows the images of the \GRB{} sky zone. The {\it rms} noise in the three images is 0.92, 0.61 and 0.28 mJy beam$^{-1}$, respectively (Table \ref{tab:image}). As a comparison, the image noise derived from the KaVA-only telescopes results in an increase by 28\%--45\%. The first epoch has only five telescopes, compared to seven telescopes in other two sessions, and thus has a relatively higher noise level.
The images cover a square region of the sky, with a side length of 
200 mas, centered at RA=03h38m01s.191, Dec=$-$26\degr56\arcmin46\arcsec.730 (J2000), i.e.~the same as 
the position determined from the VLA observation \cite{GCN23726}. The positional uncertainty given by the VLA observation is 0.04\arcsec{} in RA direction and 0.02\arcsec{} in Dec direction, respectively. Therefore a field with a radius of 100 mas is large enough to search for the GRB signal. An 
integration time of 1 second was used in correlation, to avoid
time smearing effects. There is no significant peak brighter than $5\sigma$ 
in either image. Figure  \ref{fig:GRB}-a displays ripples in which there are some bright points with a maximum brightness
of $\sim4\sigma$. The ripples are caused by 
 sparse sampling of the visibilities 
resulted from a limited number of telescopes (five, corresponding to 10 geographical baselines). These $4\sigma$ points do not have corresponding counterparts in Figure \ref{fig:GRB}-b and Figure \ref{fig:GRB}-c, whose noise level is 1.5--3 times lower. Therefore we regard these peaks as noise. 
In contrast, Figures \ref{fig:GRB}-b and -c show a smooth field without any clues of dominant signals which can be identified as candidate sources. 
In conclusion, we did not detect \GRB{} from the EAVN data in the three epochs. A $3\sigma$ upper limit is given as 2.76 mJy (2019 January 21), 1.82 mJy (2019 January 30) and 0.84 mJy (2019 February 16) respectively and used to constrain the model parameters below.

The afterglow of GRB 190114C has been studied in a number of recent works, some of which are still in preprint form, while others have been published. Wang et al. 2019 \cite{Wang2019} model the multi-wavelength emission after $\sim$1ks as synchrotron radiation from the forward shock. A similar modelling is proposed by Fraija et al. \cite{Fraija2019}, where the early optical/X-ray emission is instead interpreted as due to the deceleration of the jet in a wind-like stratified medium. The radio (1.3-97 GHz) emission in the early phase up to $\sim$1ks has been interpreted as due to the emission by the transient reverse shock that marks the time during which the jet injects its energy into the shocked region \cite{Laskar2019}. At present two possible scenarios seem to emerge in the interpretation of the available multi-wavelength data: either emission from the reverse/forward shock component in a wind medium (where the reverse and forward shock emission dominate the early- and late-time emission, respectively) or emission dominated solely by the forward shock produced by the expansion of the outflow in a medium whose density evolves from a wind environment to an interstellar medium (ISM) with a constant number density \cite{Fraija2019}. 

These different scenarios will be best tested by considering all the available multi-wavelength data sets. In particular the TeV data by MAGIC can constrain the fraction of energy of the electrons $\epsilon_{e}$ along with other afterglow parameters (see e.g. \cite{Derishev2019}). Here we show to what extent our three radio observations can constrain the parameter space of a standard forward shock synchrotron afterglow model (given that, at our observation time, any reverse shock contribution would not be relevant anymore). We consider the two possible external medium scenarios (ISM or wind) and three possible values of the parameter $\epsilon_{e}=[0.01,0.1,0.9]$ which correspond to the range of values adopted to model this afterglow in the literature. The afterglow flux is governed, among other parameters, by the outflow kinetic energy: we consider two possible values $E_{\rm k}=[10^{53},10^{54}]$ erg. Considering the isotropic equivalent energy radiated in $\gamma$--rays  of $\sim3\times 10^{53}$ erg as measured from the prompt emission spectrum (e.g. \cite{2019A&A...626A..12R}), the assumed values for $E_{\rm K}$ correspond to an efficiency of the prompt phase ($\eta\sim E_{\gamma,\rm iso}/E_{\rm k}$) of 75\% and 30\%, respectively. Finally, we assume a typical spectral index of the electron power-law distribution $p=2.3$ (this parameter has little impact on the emission at our frequencies). With these assumptions, we explore the parameter space given by the combination of the energy in the magnetic field, described through parameter $\epsilon_{\rm B}$, and the density of the external medium $n$ (or the normalization of the wind profile $A_\star$) and require the afterglow emission at 6, 15 and 32 days at 22 GHz to be consistent with our upper limits. In Fig.~\ref{fig:lightcurves} we show the parameter space allowed by our three upper limits with colored regions for the possible combinations of the assumed parameter values described above. Each shade of colour refers to a different value of $\epsilon_\mathrm{e}$; the two rows refer to the two considered values of isotropic-equivalent kinetic energy $E_\mathrm{k}$; the two columns refer to the wind (left-hand column) and uniform ISM (right-hand column) cases. Additionally, in Fig.~\ref{fig:lightcurve_comparison} we compare our limits with the expected 22 GHz emission for the various parameter sets proposed so far in the literature. Only the parameter values of Wang et al. \cite{Wang2019} are in clear tension with our upper limits.

\section{Summary}
\label{sect:discussion}

We carried out follow-up observations of \GRB{} with the EAVN at 22 GHz in three epochs, on 6, 15 and 32 days after the burst respectively. In none of the images derived from the three VLBI observations we find a source with brightness above $5\sigma$, suggesting that the source flux density has rapidly declined. 
The present observations set upper limits of radio emission of \GRB, offering useful constraints on the parameter space (e.g., energy fraction of the magnetic field, density of external medium, kinetic energy) of the synchrotron afterglow model of a standard forward shock. The modelled light curves in the recent papers are found to be consistent with our upper limits.

In addition to its scientific 
importance, this observation campaign is of particular merit: it is the first time for the EAVN to conduct fast-response observations of a transient.  
The experience gained, {\it i.e.}, completion of the sequence of proposing the observations, 
coordinating the efforts at the various stations, 
scheduling and 
conducting the first-epoch observation within one week after the X-ray triggering, is very useful for a 
future Target-of-Opportunity (ToO) operational mode of the EAVN.
Moreover, this is the first phase-referencing observation of the EAVN. So far, the EAVN has mostly been limited to the observations of bright sources (e.g., \cite{2016PASJ...68...77A,2017PASJ...69...71H,2019MNRAS.486.2412L}), therefore the capability of phase referencing 
will greatly expand its science applications.

\section{Acknowledgments}
We wish to thank the anonymous referees for their constructive comments that enabled us to improve the manuscript.
We acknowledge the financial support by the National Key R\&D Programme of China (grant number 2018YFA0404603), PRIN-INAF 2016.
T.A. thanks the grant support from the Youth Innovation Promotion Association of CAS.
We are grateful to the EAVN Directors, the scheduler, and the staff at the stations and at the KJCC for making the observations into reality.
We thank Lara Nava and Maria Edvige Ravasio for their helpful discussion and constructive comments on the manuscript.

\medskip
\medskip

{\large 

\newpage

\begin{figure}[H]
\centering
   \includegraphics[width=0.32\textwidth]{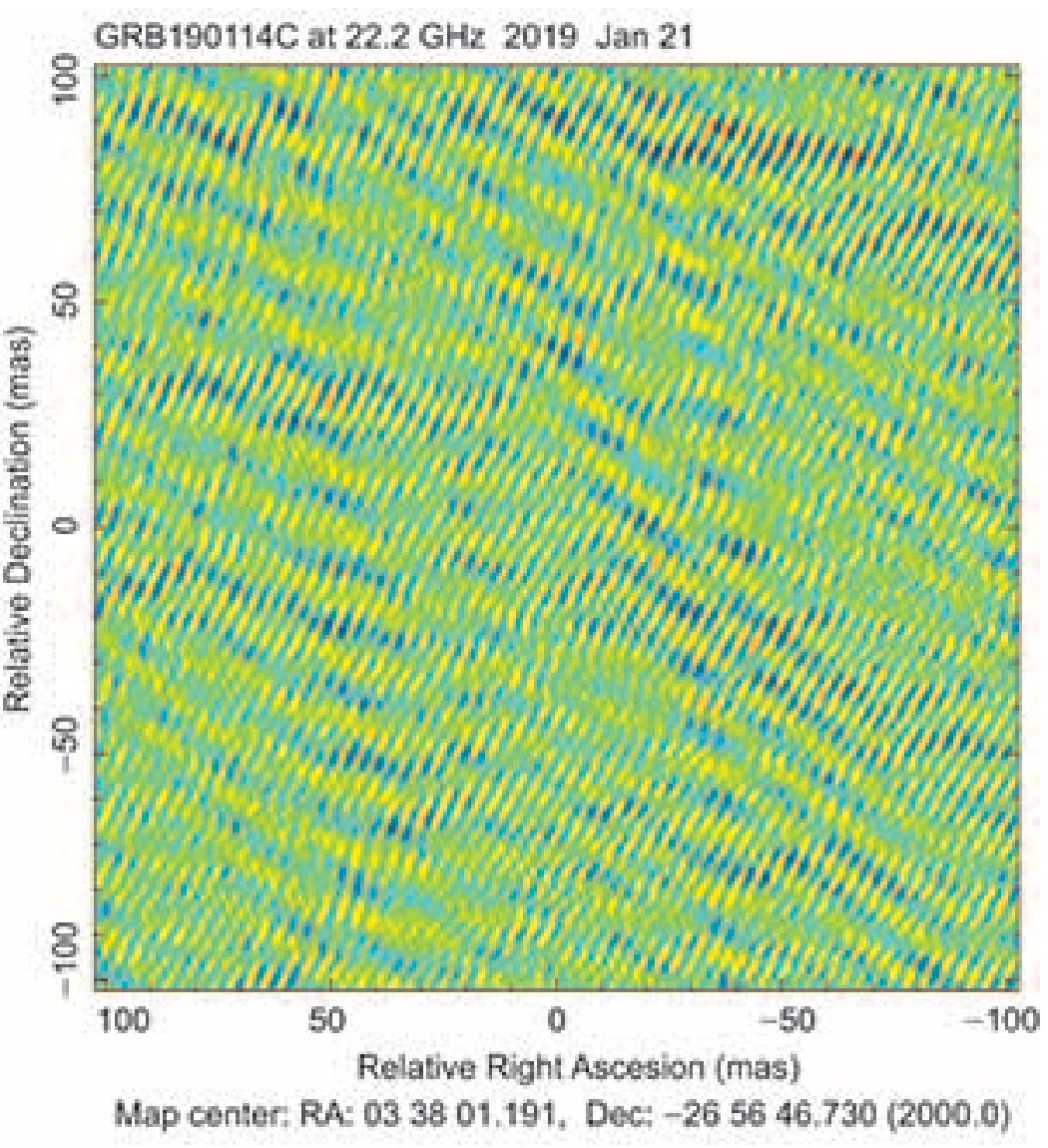}
   \includegraphics[width=0.32\textwidth]{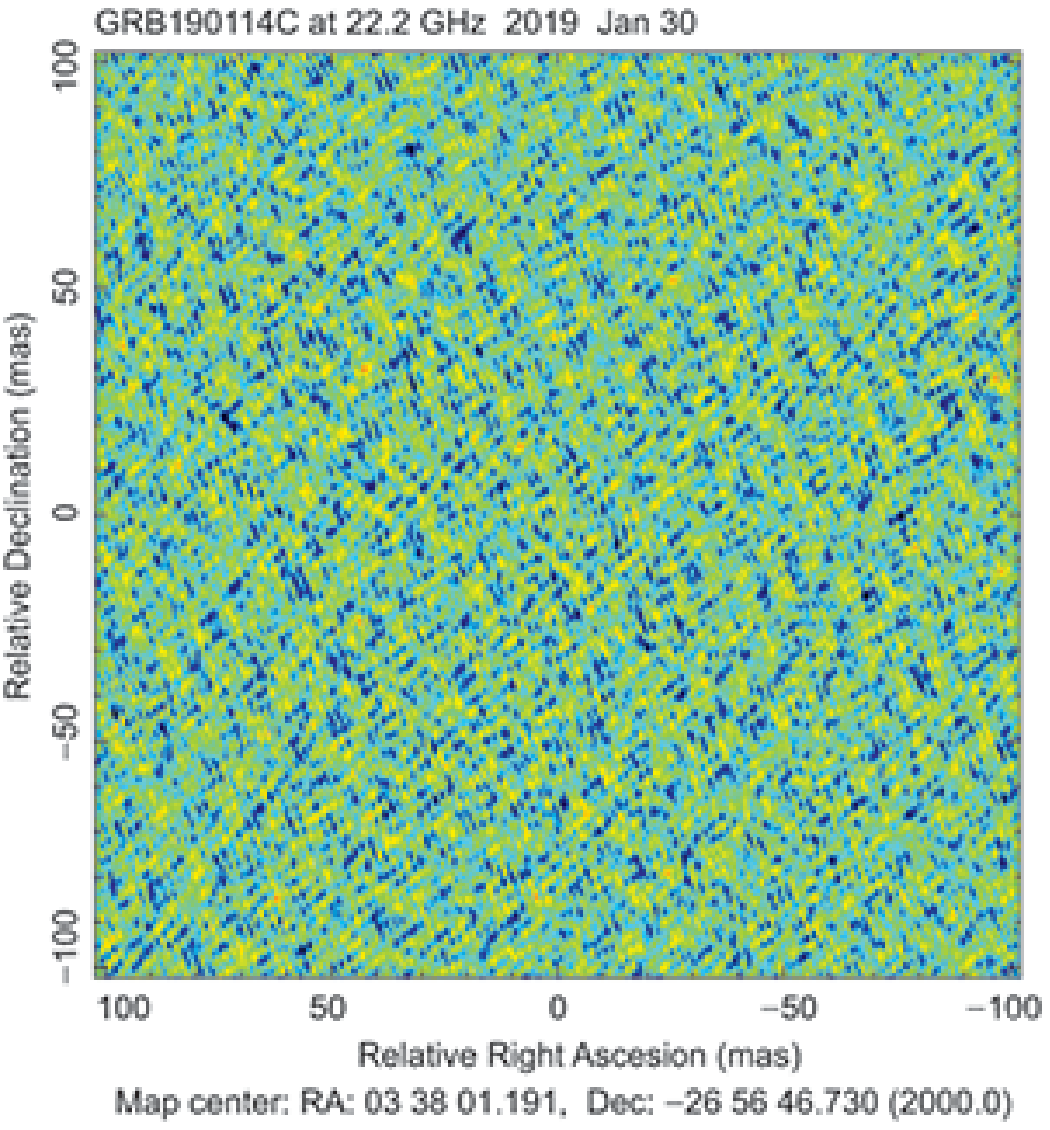}
   \includegraphics[width=0.32\textwidth]{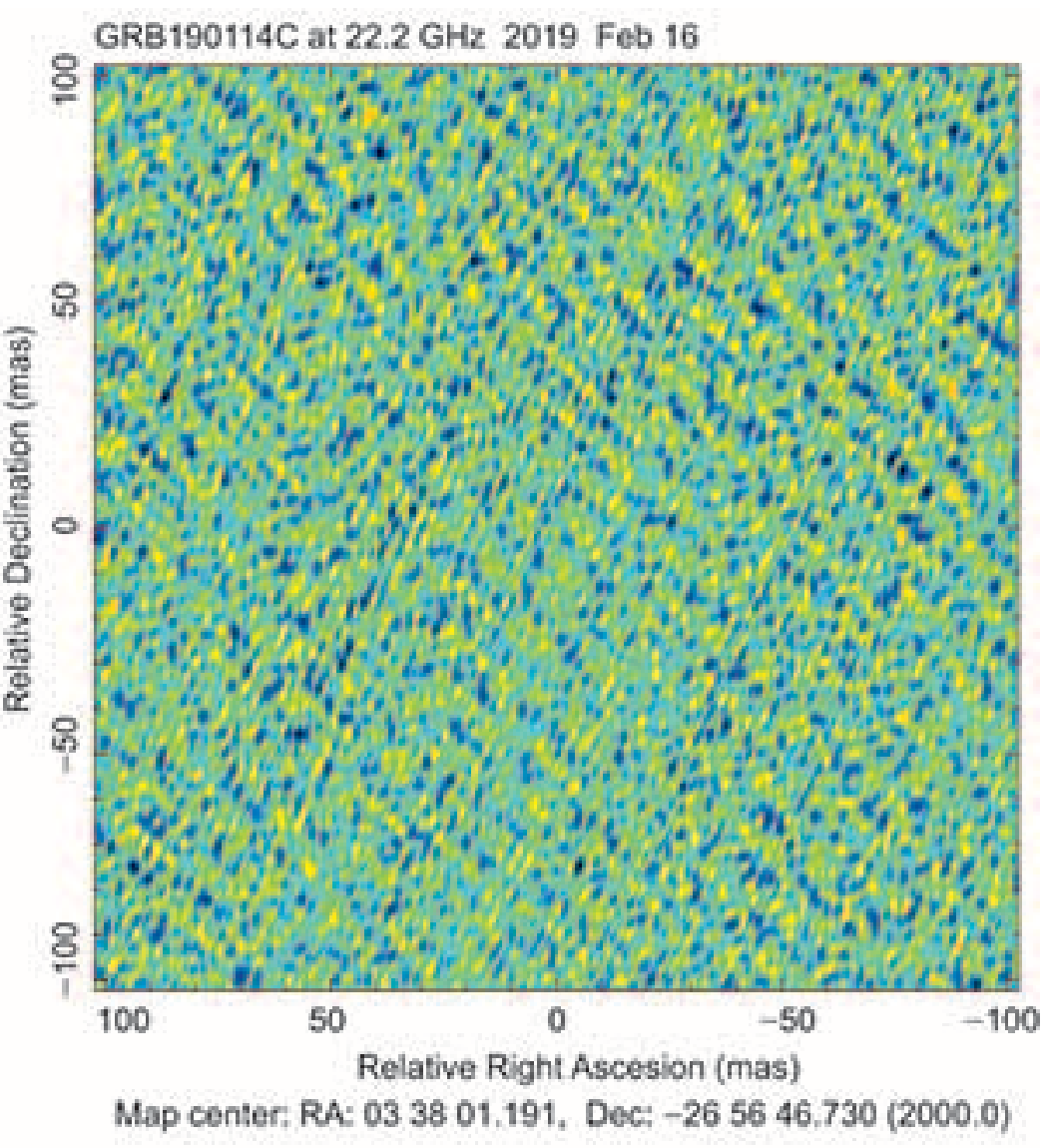}
\caption{The EAVN images of the target \GRB. The images are made by using natural weighting. The color scale represents the brightness: from $-$3.69 (dark blue color) to 4.24 (red color) mJy beam$^{-1}$ (panel a), from $-$2.7 to 2.67 mJy beam$^{-1}$ (panel b), from $-$1.19 to 1.18 mJy beam$^{-1}$ (panel c). The rms noise is 0.9 mJy beam$^{-1}$, 0.6 mJy beam$^{-1}$, and 0.3 mJy beam$^{-1}$ for panels a, b and c, respectively. The image center (RA=03h38m01s.191, Dec=$-$26\degr56\arcmin46\arcsec.730) was adopted from the VLA observation \cite{GCN23726}. There is no significant source brighter than 5$\sigma$ in the 200 mas $\times$ 200 mas field. }
\label{fig:GRB}
\end{figure}

\begin{figure}[H]
    \centering
    \includegraphics[width=0.95\columnwidth]{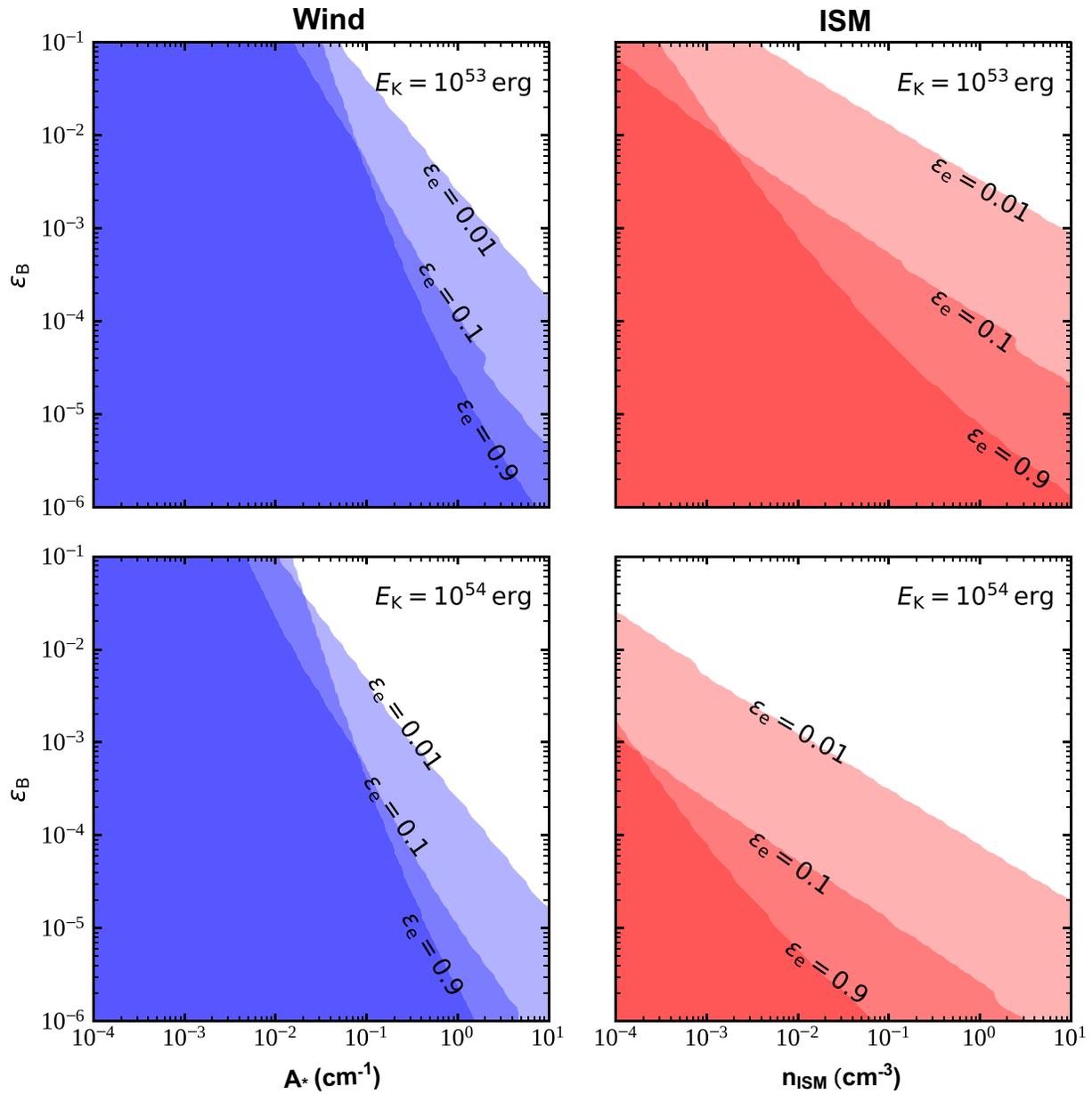}
    \caption{Constraints on the forward shock synchrontron afterglow model parameters. The panels show either the 2D parameter space $\epsilon_{
    \rm B}$--$A_*$ (for the wind external medium -- left-hand column) or $\epsilon_\mathrm{B}$--$n_{\rm ISM}$ (for the uniform ISM external medium -- right-hand column). The parameter $A_{\star}$ is the density parameter for the wind medium (as defined in \cite{Chevalier2000}). Coloured regions correspond to parameters that produce radio emission consistent with our upper limits. Each plot shows three shaded regions corresponding to three different values of $\epsilon_\mathrm{e}$, namely $0.01$, $0.1$ and $0.9$. The two rows correspond to two possible values of the outflow kinetic energy, namely $E_\mathrm{K}=10^{53}\,\mathrm{erg}$ and $10^{54}\,\mathrm{erg}$. }
    \label{fig:lightcurves}
\end{figure}

\begin{figure}
    \centering
    \includegraphics[width=0.95\columnwidth]{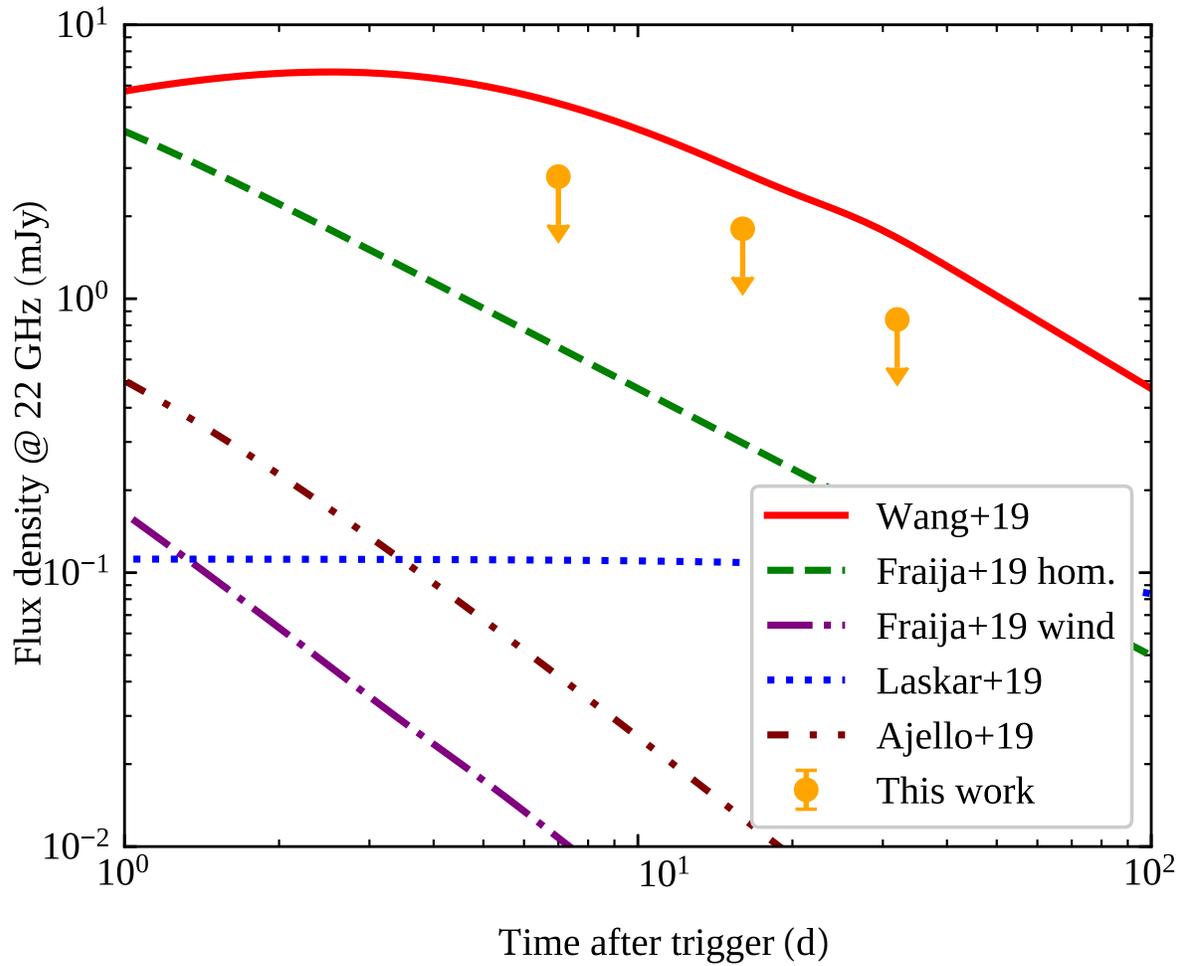}
    \caption{Comparison of our upper limits with the expected 22 GHz afterglow light curves for different models proposed in the literature. The three orange circles with downward arrows represent our three-sigma upper limits. Each line represents the expected 22 GHz afterglow light curve for a different set of parameters, computed following \cite{Granot2002}. The sets of parameters correspond to those used in a number of recent papers that model the GRB 190114C afterglow, as reported in the legend. The parameters of \cite{Wang2019} are excluded by our observations.}
    \label{fig:lightcurve_comparison}
\end{figure}

\label{lastpage}

\newpage

\begin{appendix}                  %%appendicial material is supported

\section{Observations and Data Reduction}

Table \ref{tab:obs} presents the observation logs.
Altogether, 
eight telescopes of the EAVN participated in the observations: Tianma 65m (TIA) of China; Iriki 20m (IRK), Ishigakijima 20m (ISG), 
Mizusawa 20m (MIZ) and Ogasawara 20m(OGA) of the VLBI Exploration of Radio Astrometry (VERA) array, Japan \cite{VERA}; Tamna 21m (KTN), Ulsan 21m (KUS), Yonsei 21m (KYS) of the Korean VLBI Network \cite{KVN}. 
Due to maintenance and/or technical problems, some telescopes did not participate in certain sessions. 
The observations were executed in phase referencing mode, involving fast nodding of the telescopes between the target and nearby bright calibrator \PRcal{} (1.2 Jy at 22 GHz, Figure \ref{fig:pha-cal}; 2.5$^\circ$ away from the target on the plane of the sky). In order to keep the visibility phase from being affected by the atmospheric fluctuations at 22 GHz, a cycle of 'calibrator (50s) -- slewing (10s) -- target (50s)' was adopted. The total time 
spent on the target at each epoch is $\sim$130 min.
Figure \ref{fig:uvcov} shows an example of the (u,v) coverage.

We should note, 
since this is the first EAVN phase-referencing observation, that 
there is no a priori experience on 
the fast switching schedule at 22 GHz. It turned out that the 10s slewing time between the target and phase-referencing calibrator is not long enough for the big antenna Tianma. Moreover, 
the system temperature of Tianma was 
abnormally high, that can not be solely due to the weather condition. The problem is possibly associated with the 22-GHz pointing model and the surface model which is adjusted by actuators. We used VERA and KVN telescopes to calibrate the visibility amplitude of Tianma.
Surprisingly, KVN antennas were also found to suffer from the slewing problem; the reason is unknown yet and to be investigated. As a result, the system temperatures of KVN telescopes in the first $\sim$20 seconds are significantly higher than the rest of the scan. We had 
to delete the 
beginning
24 second ($\sim50\%$) visibility data at each scan. This significantly reduced the effective on-source time of KVN telescopes. The data loss due to antenna slewing and absence of certain telescope(s), as well as the high atmospheric opacity at 22 GHz, results in a decrease of the sensitivity. The actual image noise is 3--7 times higher than 
the ideal sensitivity
calculated with the antenna's system equivalent flux density (SEFD) reported in the EAVN website, a parameter that measures the overall performance of each antenna (usually measured at the good operational condition).

The raw data were recorded in eight  
subbands
each 32 MHz using left-handed circular polarization; 2-bit quantization sampling was adopted. These setups resulted in a recording data rate of 1 Gbps 
at each station. All the data  observed at the stations were recorded in disk modules which were shipped to the Korea-Japan Correlation Center (KJCC) at Daejeon, Korea, where the correlation was processed. The correlation integration time was 1 s, corresponding to a largest field of view (FoV) without being affected by time smearing of about 4.8\arcsec.

The data reduction and calibration of the observed visibility amplitude, phase and bandpass were performed using the US National Radio Astronomy Observatory (NRAO) Astronomical Image Processing System (AIPS) \cite{2003ASSL..285..109G} based on the standard EAVN data reduction procedures (e.g., \cite{2017MNRAS.468...69Z,2017PASJ...69...71H,2019MNRAS.486.2412L}).
A priori amplitude calibration (except for Tianma, see above discussion) was done using the system temperatures measured during the observations and the antenna gain curves supplied by the EAVN stations.
A bright calibrator source 0420$-$014 was used to calibrate the instrumental phase offsets between the subbands. After that the global fringe fitting for the phase referencing calibrator \PRcal{} were performed. The calibrated visibility data of \PRcal{} were exported into the 
{\sc Difmap} software package \cite{1994BAAS...26..987S} for imaging and self-calibration.
Figure \ref{fig:pha-cal} shows the {\sc CLEAN} image of \PRcal{} after several iterations of hybrid imaging loops. It displays a core-jet structure.
The CLEAN 
model components
 of \PRcal{} were imported back into {\sc AIPS}.
We run fringe fitting again by using these 
CLEAN model components
which represent the core-jet structure.
The derived phase, delay and rate solutions from \PRcal{} were applied to \GRB{} data by interpolating.
The data were finally averaged in frequency within each subband, but individual subbands were kept to minimize bandwidth smearing. Similarly, the data were time-averaged to 2 seconds, which allows for searching the source within a FoV of 2.4\arcsec, as the position uncertainty determined by the VLA is 0.02\arcsec--0.04\arcsec.

\begin{table}[H]
\caption{Observation log.}
\begin{tabular}{ccccc}
  \hline\noalign{\smallskip}
Epoch & Telescopes & BW &  Time Range & Beam (maj, min, PA)  \\
%       &       & (MHz) & (hour) & (mas, mas, $^\circ$) \\
(1) & (2) & (3) & (4) & (5) \\
  \hline\noalign{\smallskip}
2019 01 21 &  TIA, IRK, ISG, OGA,  KUS          & 256 MHz & UT 08:30--14:30 & 2.17 mas $\times$ 1.08 mas, 2.4$^\circ$ \\
2019 01 30 & TIA, IRK, ISG, OGA, KTN, KUS, KYS  & 256 MHz & UT 08:00--14:00 & 2.25 mas $\times$ 1.12 mas, 11.1$^\circ$ \\
2019 02 16 & TIA, IRK, ISG, MIZ,  KTN, KUS, KYS & 256 MHz & UT 08:00--14:05 & 3.38 mas $\times$ 1.44 mas, $-$13.4$^\circ$ \\
  \hline\noalign{\smallskip}
\end{tabular} \\
Note: (2) a core array of the EAVN consisting of the participating telescopes: Tianma 65m (TIA) of China;  Iriki 20m (IRK), Ishigakijima 20m (ISG), Mizusawa 20m (MIZ) and Ogasawara 20m (OGA) of the VERA, Japan; Tamna 21m (KTN), Ulsan 21m (KUS), Yonsei 21m (KYS) of the KVN, Korea. KVN and VERA are often combined to the KaVA subarray.
\label{tab:obs}
\end{table}

\begin{figure}[H]
\centering
  \includegraphics[width=0.7\textwidth]{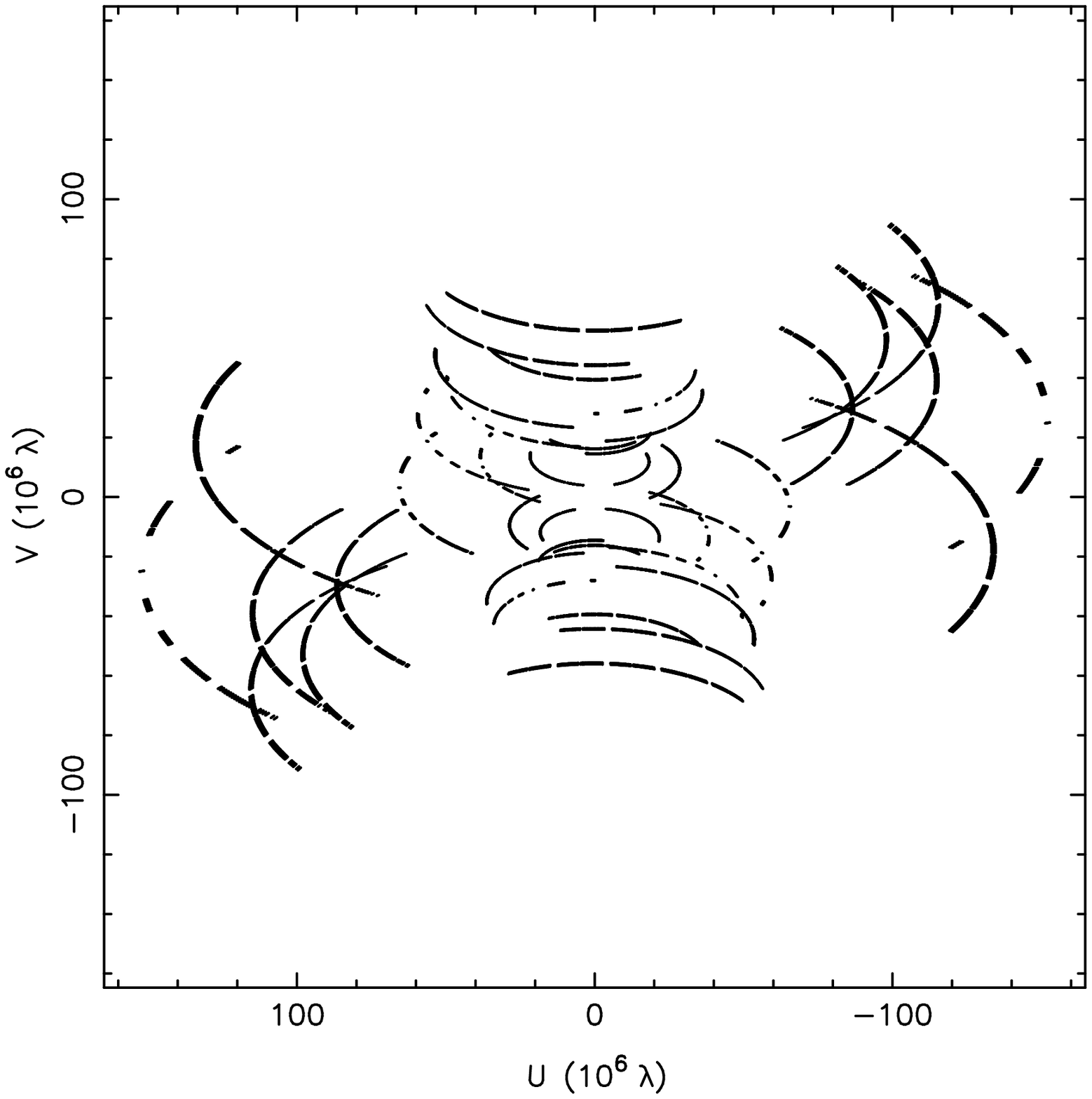}
\caption{The (u,v) coverage of the EAVN observation made on 2019 January 30.}
\label{fig:uvcov}
\end{figure}

\section{Images of calibrator}

Figure~\ref{fig:pha-cal} shows the core-jet structure of the phase-referencing calibrator \PRcal{} observed on 2019 February 16. The images on the
other two epochs are similar in 
structure, thus are not shown here. The effective on-source time is $\sim$2 hours, resulting in an rms noise of 0.6 mJy beam$^{-1}$. As a comparison, the latest observations of M87 with KVN and VERA (KaVA)--only Park et al. 2019 \cite{Park2019} presents an image noise of 0.3--0.6 mJy beam$^{-1}$, representing the typical performance of the KaVA. Both sources, \PRcal{} in this paper and M87 in \cite{Park2019}, are $\sim$1 Jy 
bright AGNs. Note that the observations of Park et al. \cite{Park2019} are not in phase referencing mode. The total on-source time in their M87 observations is $\sim$4.5 hours, two times longer than ours, but our observations contain the large 
telescope Tianma 65m. When using KaVA--only data, the derived image noise of the \PRcal{} image in the third epoch is 0.8 mJy beam$^{-1}$ (Table \ref{tab:image}). Then the sensitivity of our observations is consistent with the typical KaVA value, showing that our observations of the calibrator are in good condition and the data were properly calibrated. \PRcal{} has rich jet structure extending from the compact core to the southeast at a projected distance of $\sim$6 mas at 22 GHz (Figure~\ref{fig:pha-cal}). It is also a prominent $\gamma$-ray source at $z = 0.991$, as 
observed by the 
Fermi/LAT
(e.g., \cite{2010ApJ...715..429A}). The source was recently detected with a strong GeV flare \cite{2018ATel11644....1C}, and multi-band observations were 
triggered (e.g., \cite{2018ATel12336....1N}). 

After applying the gain solutions derived from \PRcal{} to \GRB, we created the images of \GRB, as are shown in Figure \ref{fig:GRB}. 
The {\it rms} noise in each image is listed in Table \ref{tab:image}.
There is no significant source which can be regarded as a detection. 
We have also tried imaging with the combined data of 
Sessions 2 and 3, 
which are only separated by 17 days. The inferred image rms 
noise is 0.33 mJy beam$^{-1}$, even higher than that of Session 3. A possible reason for the increased 
noise of the combined data is that misalignment of the visibility amplitude owing to the uncertainty of the amplitude scales (at a level of 5\%--10\%) adds to the contribution of the noise budget.

\begin{figure}[H]
\centering
  \includegraphics[width=0.8\textwidth]{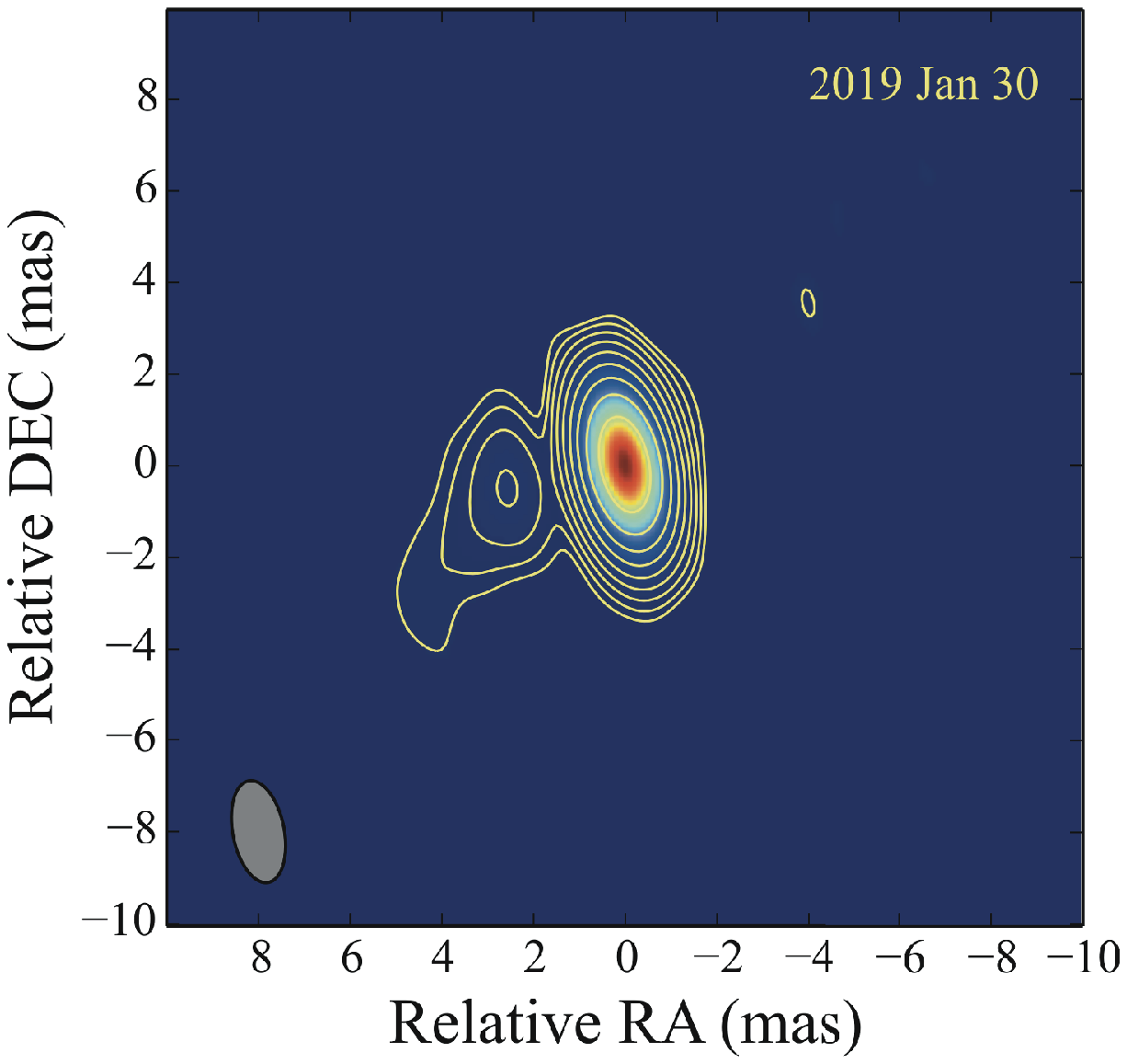}
\caption{Total intensity image of the calibrator J0348$-$2749 showing a core-jet structure. The image is made by using natural weighting. The color scale represents the intensity, ranging from 0.3 mJy beam$^{-1}$ (blue color) to 1 Jy beam$^{-1}$ (red color). The peak brightness is 1.01 Jy beam$^{-1}$.  The rms noise is 0.7 mJy beam$^{-1}$. The lowest contour represents three times rms noise level. The contours increase in steps of 2.  The restoring beam, the grey-colored ellipse in the left-bottom corner, is 2.3 mas $\times$ 1.1 mas, at a position angle of $11.1^\circ$.}
\label{fig:pha-cal}
\end{figure}

\begin{table}[H]
\caption{Image noise (in unit of mJy beam$^{-1}$) at different epochs and with different telescopes.}
\begin{tabular}{c|cc|cc}
  \hline\noalign{\smallskip}
Epoch &  EAVN$^a$& KVN+VERA &  EAVN & KVN+VERA \\
      & \PRcal   & \PRcal   & \GRB  & \GRB     \\
(1)   & (2)      & (3)      & (4)   & (5)      \\
  \hline\noalign{\smallskip}
Session 1 (2019 01 21) & 1.15 & 1.27 & 0.92 & 1.32 \\
Session 2 (2019 01 30) & 0.72 & 0.88 & 0.61 & 0.78\\
Session 3 (2019 02 16) & 0.59 & 0.80 & 0.28 & 0.39 \\
  \hline\noalign{\smallskip}
\end{tabular} \\
$^a$: the EAVN array in these observations is comprised of the three telescopes of the Korean VLBI Network (KVN), four telescopes of the VLBI Exploration of Radio Astrometry (VERA), and Tianma of China.
\label{tab:image}
\end{table}

\end{appendix}
\medskip
\medskip
\textbf{Appendix C} \\

% \end{multicols}

\end{document}